# DESIGN OF THE MPRI CONTROL SYSTEM

J.C. Collins, M. Ball, B. Broderick, J. Katuin, Wm. Manwaring, N. Schreuder
IUCF, Bloomington, IN 47408, USA

Abstract

The Indiana University Cyclotron Facility (IUCF) is in the process of building the Midwest Proton Radiation Institute (MPRI). This involves refurbishing the 200MeV cyclotron and building new facilities for the purpose of providing clinical treatment of human cancer tumors. First patients are expected in the Spring of 2003. This paper presents the design and implementation to date of the controls, hardware and software, for both accelerator and treatment areas. Particular attention is placed on issues of personnel safety and control system security, development of inexpensive VMEbus boards with emphasis on the use of MicroChip PIC processors, beam diagnostics and monitoring and the use of commercial robots and vision systems for patient positioning.

## 1 INTRODUCTION

Faced with the end of support for physics research on its 200MeV cyclotron, IUCF has chosen to pursue further use of the accelerator for the treatment of various cancer tumors. To this end, the cyclotron system is being modified to produce protons at the single energy of 205MeV, the cyclotron experimental hall has been cleared and a trunk beam line with separate treatment beam lines are being installed.[1]

Controls for both the cyclotron and all beam lines will use an architecture very similar to that used in the most recent IUCF projects[2]. Two minicomputers, one for the cyclotron, one for all treatment beam lines, provide the user interface and all related processing. Each uses a PCI-VME interface to communicate with a master VMEbus crate, which is connected to geographically distributed crates via fiber optic bus extenders. The minicomputers run OpenVMS and Vsystem. On/Off controls and interlocking are done with PLCs, using centralized processors and distributed I/O modules.

The full plan calls for three treatment rooms, two of which are to hold gantries. Each treatment room will have its own, independent room control and dose monitoring computers. In all rooms, accurate and reproducible patient positioning will be achieved using commercial robots. But because robot motion has more degrees of freedom than needed, an independent computer vision system will be used to track and interlock robot motion. Finally, all patient treatment parameters will be supplied to the various control computers from a medical computer system.

## 2 HARDWARE

### 2.1 Beam Line Control

The cyclotron and beam line control computers are Compaq DS10 AlphaStations, each with an SBS Bit3 PCI-VME interface to a 6U VME crate. In turn, this crate contains VMIC 5531L modules which connect to remote 6U and 3U VME crates via fiber optics.

For DACs and ADCs, the standard debate between buying commercial or designing in-house was decided in favor of the latter option. Using FPGAs and out-sourcing construction allows even a small facility to do such a project. This option is not more expensive than the commercial one when requirements for good galvanic isolation and a high degree of modularity (for quick repair) are included.

In fact, we have developed four single width VME boards:

- DAB (Data Acquisition Board) - 3U board contains one DAC and six ADC channels.
- SDPM (Analog Dual Port Memory) - 6U board connects to up to eight PIC processors, each connected to a serial (e.g., RS-232) device.
- ADPM (Serial Dual Port Memory) - 6U board connects to up to eight PIC processors, each having four channels of 12-bit ADC.
- DIO (Digital Input and Output) - 3U board with two 16-bit channels; supports VMEbus interrupts.

Interlocked functions using digital I/O are performed with an Allen-Bradley PLC and Flex-IO modules.

### 2.2 DAB

All DAB channels are 2's complement, 16 bits and ± 10V. The analog section of the board is optically isolated from the VME power, ground and signal bus. The six ADC channels are paired internally, each pair sharing an external connection. All channels have low-pass filters, but with the response of one (slow) channel in each pair having greater attenuation with frequency than the other. The slow channel will be quiet and used for normal operator displays, while the other is available to diagnostic software looking for high frequency components signaling problems.

## 2.3 SDPM, ADPM

These boards are variants of the same basic design. Originally conceived to be an inexpensive interface to devices with slow, serial interfaces, the boards present two banks of 1KB to the host computer. While the PIC processors gather data and store it in one memory bank, the host computer can read data from the other. Bank switching is under host control. Adding 12-bit, 2's complement ADCs to the remote PICs created an inexpensive method of gathering slow analog data close to the beam line. The RS232 or analog section of each board is optically isolated from the VME power, ground and signal bus. Figure 1 shows how beam line elements are connected to the control system using DAB and ADPM.

Figure1. Beam Line Element Connection.

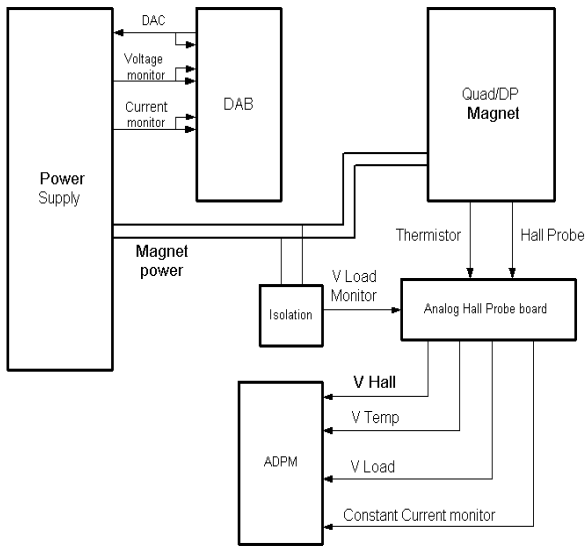

## 2.3 Timing

The few timing signals required will be supplied by a single sequencer module[3]. Depending on how it is configured, the sequencer can provide up to 32 signal outputs with 1ìsec resolution or DAC output synchronized with timing signals. The latter setup may control beam intensity when using multiple treatment rooms.

# 3 SOFTWARE

## 3.1 Overview

The control computers will run OpenVMS and Vsystem, a combination IUCF has used for many years. Each treatment room control computers will run a version of Windows because of the need to run commercial software. The dose monitoring computer will run a small, realtime OS which has not been selected yet. All these computers will have Ethernet connections and be separated from the rest of IUCF by managed switches and from the rest of the world by firewalls.

## 3.2 Cyclotron Control

In December, 2000, the PDP-11/44 which had run the cyclotron since 1985 was replaced with a DS10. An interface between VME and the original DIO control bus was installed and about 20% of the control hardware was replaced before the cyclotrons were brought back into operation using Vsystem in March, 2001. We will continue to replace portions of the old cyclotron controls as time and access permit.

## 3.3. Beam Line Control

Basic beam line control will be based on programs previously created and tested at IUCF. Major changes or additions to that software include:

- 2-of-3 Checking – As shown in Figure 1, each beam line element will have power supply current, load voltage and a temperature corrected Hall probe magnetic field readout. All three values must be within acceptable ranges before operation is allowed. There will be a procedure whereby the operator will be able to specify that operation is allowed with one out-of-range value for a short, fixed time (until the next maintenance period).
- Machine state restore – There will be one "golden" set of DAC and ADC values defining proper accelerator and beam line operation. That will be the only set supported for computer restoration.
- Energy setup – A degrader in each treatment beam line will be used to change the fixed cyclotron output proton beam energy to that desired for a specific treatment. We will set each treatment line element by interpolating in a table of empirical values and check that beam transmits through the line as expected before allowing a real treatment to start.
- Security – No remote logins will be permitted for the treatment room controllers. Restrictions will

be placed on those network nodes from which users may log into the control system. All controls, except a beam stop and closed loops, will be disabled while a treatment is underway.

## 4 DIAGNOSTICS

### 4.1 Beam Position Monitor (BPM)

BPMs[4], non-intercepting transducers that provide both position and intensity information, are the primary beam tune-up and monitoring tool. The BPM system has four major sections, the pickup, the radio frequency (rf) amplifiers and mixers, the analog and conversion section, the analog-to-digital converter (ADC), and the software. They provide ±0.2mm position resolution over a ±12mm area and 10% intensity accuracy for beams from 10nA to 2ìA. Operator displays are typically a set of bar graphs showing all readouts for a single beam line.

### 4.2 Harp

A harp is a multi-wire device which can measure the beam profile in one transverse plane using either secondary emission electrons or ionization (gas harp). Our harps will use a wire spacing of 0.5mm and operate between 10nA and 2ìA beam current. Harps are usually mounted in X,Y pairs and operators are shown data from both as wave forms.

### 4.3 Wire Scanner

This is a wire helix inserted in the beamline at 45º and rotated so as to cross the beam once in each transverse plane on each rotation. Wire scanner enable and selection will be done with an in-house built controller and multiplexer which enables an operator to select any two wire scanners. Operators are shown X,Y results for the selected wire scanner(s).

### 4.4 Multi-Layer Faraday Cup (MLFC)

Used to characterize the energy and energy distribution of the beam, the MLFC consists of a solid block followed by insulated plates[5]. The set of currents read from the plates can be used to determine if the beam energy properties are correct. The MLFC will be able to resolve energy drifts >100KeV and energy spread changes >500KeV using currents between 10 and 500nA.

## 5 PATIENT POSITIONING

For treatment, the patient (actually, the tumor) must be accurately and reproducibly positioned with respect to the proton therapy beam line. This will be accomplished using a Motoman UP200 robot and XRC controller. This robot provides six degrees of freedom, a rated payload of 200kg and positioning accuracy of ±0.2mm throughout its 2446mm working envelop. A precalculated and tested (simulated) robot job file defining all patient-specific robot motions will be part of the patient treatment prescription obtained from medical records when the patient is identified. Local, manual control of the robot will be provided through a wireless, handheld PC.

A robot is almost perfect for the patient positioning job except that its motion has six degrees of freedom, i.e., it is capable of turning the patient upside down. To stop such undesired movements, the patient support device (bed or chair) will be fitted with an acceleration and tilt monitoring system independent of the robot controller. The patient support device will also have special markers attached which will be watched by a Vicon 460 Vision System. The Vision System uses infrared to track each marker and report its position many times per second. Knowing the expected robot path and the fixed geometry of the room, we can determine if the robot is doing something unexpected or about to hit an object and turn the robot off.

## REFERENCES

[1] D. Friesel, et al., "The Indiana University Midwest Proton Radiation Institute", PAC2001, Chicago, IL, June 2001.

[2] J.C. Collins, et al., "The CIS Control System at IUCF", PAC1999, New York, NY, March 1999.

[3] W. Hunt, "Cooler Injector Synchrotron Control Hardware at IUCF", PAC1999, New York, NY, March 1999.

[4] M. Ball, et al., "Beam Diagnostics in the IU Cooler Injector Synchrotron", 7[th] Beam Instru. Workshop, Argonne, IL, 1996.

[5] N. Schreuder, "The Development of a Multi-Layer Faraday Cup to Characterize the NAC Proton Therapy Beam", PTCOG-XXXII, Upsala, 1999.